\documentclass[12pt]{iopart}
\usepackage{epsfig}
\usepackage{iopams}
\newcommand{\eq}{\begin{eqnarray}}
\newcommand{\en}{\end{eqnarray}}
\newcommand{\la}{\langle}
\newcommand{\ra}{\rangle}
\begin{document}

\title{Dependence of nucleon properties on pseudoscalar meson masses}

\author{
Amand Faessler$^1$, 
Thomas Gutsche$^1$,
Valery E Lyubovitskij$^1$\footnote{On leave of absence from the
Department of Physics, Tomsk State University,
634050 Tomsk, Russia}, 
Chalump Oonariya$^{1,2}$} 

\address{
$^1$ Institut f\"ur Theoretische Physik, Universit\"at T\"ubingen, \\
Auf der Morgenstelle 14, D-72076 T\"ubingen, Germany\\[2mm]  
$^2$ School of Physics, Suranaree University of Technology, \\ 
111 University Avenue, Nakhon Ratchasima 30000, Thailand 
}

\ead{amand.faessler@uni-tuebingen.de,
thomas.gutsche@uni-tuebingen.de, \\
valeri.lyubovitskij@uni-tuebingen.de,
chalump@tphys.physik.uni-tuebingen.de}

\begin{abstract}
We discuss the sensitivity of nucleon properties (mass, magnetic moments  
and electromagnetic form factors) on the variation of the pseudoscalar  
meson masses in the context of the perturbative chiral quark model.  
The obtained results are compared to data and other theoretical predictions. 
\end{abstract}

\vskip 1cm


\vskip 1cm


\newpage

\section{INTRODUCTION}
In recent years nucleon properties have been in the focus of 
manifestly Lorentz covariant Chiral Perturbation Theory (ChPT), 
improved lattice QCD computations and chiral extrapolations (see e.g. 
Refs.~\cite{Becher:1999he}-\cite{Hemmert:2003ta}). 
The lattice formulation of QCD is well established and is a powerful 
tool for studying the structure of nucleons. The computation of nucleon 
properties in lattice QCD is progressing with steadily increasing 
accuracy~\cite{Procura:2006bj}-\cite{Ali Khan:2003cu}.   
Accurate computations of the nucleon mass with dynamical fermions and two
active flavors are now possible~\cite{Ali Khan:2001tx,Aoki:2002uc} 
in lattice QCD.
In practice, these computations are so far limited to relatively large
quark masses. Direct simulations of QCD for light current quark
masses, near the chiral limit, remain computationally intensive.
To extract predictions for observables, lattice data generated at high
current quark masses have to be extrapolated to the point of physical 
quark or pion mass. Therefore, one of the current aims in lattice QCD 
is to establish the quark mass dependence of quantities of physical
interest, such as the nucleon mass, magnetic moments and form factors.
The major tool in establishing the current quark mass dependence
of lattice QCD results are methods based on chiral effective field theory.
Recent extrapolation studies of lattice results concern the nucleon 
mass~\cite{Procura:2006bj,Procura:2003ig,Young:2002ib,Bernard:2003rp,%
Frink:2005ru}, its axial vector coupling constant 
and magnetic moments~\cite{Young:2002ib}-\cite{Holstein:2005db},  
the pion-nucleon sigma term, charge radii~\cite{HackettJones:2000js}, 
form factors~\cite{Hemmert:2002uh}-\cite{Wang:2007iw}, 
and moments of structure functions~\cite{Detmold:2001jb}.
The chiral expansion in chiral effective field theory 
($\chi$EFT) has been used to study the quark mass (pion mass) dependence 
of the magnetic moments, magnetic form factors and the  
axial-vector coupling constant~\cite{Hemmert:2003cb,Procura:2006zv} of the 
nucleon for extrapolations of lattice QCD results, so far determined at 
relatively large quark masses corresponding to pion masses of 
$m_\pi \ge $ 0.6 GeV, down to physical values of $m_\pi$. In the chiral limit, 
with $m_\pi \to 0$, QCD at low energies is realized in the form of an 
effective 
field theory with spontaneously broken chiral symmetry, with massless pions 
as the primary active degrees of freedom. The coupling of the chiral Goldstone 
bosons to these spin-1/2 matter field produces the so-called "pion-cloud" of 
the nucleon, an important component of nucleon structure at low energy and
low momentum scales. 

In the present paper we investigate the dependence of nucleon
properties (mass, magnetic moments and electromagnetic form factors)
on pseudoscalar meson masses applying the perturbative chiral quark 
model (PCQM)~\cite{Lyubovitskij:2001nm}-\cite{Cheedket:2002ik}.  
In the PCQM baryons are described by three relativistic
valence quarks confined in a static potential, which are supplemented by a
cloud of pseudoscalar Goldstone bosons, as required by chiral symmetry.
This simple phenomenological model has already been successfully applied 
to the charge and magnetic form factors of baryons, sigma terms, ground 
state masses of baryons, the electromagnetic $N \to \Delta$ transition, 
and other baryon properties~\cite{Lyubovitskij:2001nm}-\cite{Cheedket:2002ik}. 
Note that in Refs.~\cite{Faessler:2005gd,Faessler:2006ky} we extend this 
approach by constructing a framework which is manifestly Lorentz covariant 
and aims for consistency with ChPT.  

In this work, our strategy is as follows. First, we discuss the
nucleon properties (mass, magnetic moments and electromagnetic 
form factors) in dependence on the pion mass in the two-flavor sector. Second, 
we extend our formalism to the three-flavor sector including kaon and 
$\eta$-meson degrees of freedom with fixed masses. 
All calculations are performed at one loop. 
The chiral limit, where current quark masses approach zero with 
$\hat m$, $m_s \to 0$, is well defined. We compare the obtained quark mass
dependence of the nucleon observables to
the results of other approaches (lattice QCD, chiral extrapolations). 

The paper is organized as follows. In Sec. II we give a short overview 
of our approach. In Sec. III we the discuss dependence of nucleon properties 
on the variation of the pion mass in the two- and three-flavor picture 
in the context of the PCQM and compare them to other theoretical 
approaches. In Sec. IV we give our conclusions. 

\section{THE PERTURBATIVE CHIRAL QUARK MODEL}

The perturbative chiral quark 
model~\cite{Lyubovitskij:2001nm}-\cite{Cheedket:2002ik} 
is based on an effective chiral 
Lagrangian describing baryons by a core of three valence quarks, 
moving in a central Dirac field with $V_{\rm eff}(r)=S(r)+\gamma^0V(r)$, 
where $r=\mid\vec{x}\mid$. In order to respect chiral symmetry, 
a cloud of Goldstone bosons ($\pi$, $K$ and $\eta$) is included, which
are treated as small
fluctuations around the three-quark core. The model Lagrangian is 
\eq\label{L_chir}
{\cal L}(x)&=&\bar\psi(x)\,[i \not\!\partial\,-\,\gamma^0 V(r) 
- {\cal M} ]\,\psi(x)\nonumber\\ 
&+&\frac{F^2}{4} \, {\rm Tr}\biggl[\partial^\mu U(x) \, 
\partial^\mu U^\dagger(x) + 2 {\cal M} B ( U(x) + U^\dagger(x)) \biggr]
\nonumber\\ 
&-& \bar\psi(x) \, S(r) \, \biggl[ \frac{U(x) \, + U^\dagger(x)}{2} \, + 
\, \gamma^5 \, \frac{U(x) \, -  U^\dagger(x)}{2} \biggr] \, \psi(x) \; ,  
\en 
where $\psi = (u, d, s)$ is the triplet of quark fields, 
$U = \exp[i\hat{\Phi}/F]$ is the chiral field in the exponentional 
parametrization, $F=88$ MeV is the pion decay constant in the chiral 
limit~\cite{Gasser:1987rb}, ${\cal M}={\rm diag}\{m_u,m_d,m_s\}$
is the mass matrix of current quarks and 
$B=-\langle 0|\bar u u|0 \rangle / F^2 = 
-\langle 0|\bar d d|0 \rangle / F^2$ is the quark condensate constant. 
In the numerical calculations we restrict to the isospin symmetry limit 
$m_u=m_d=\hat m$. We rely on the standard picture of chiral symmetry
breaking~\cite{Gasser:1982ap}  and for the masses of pseudoscalar mesons
we use the leading term in their chiral expansion (i.e. linear in the
current quark mass). By construction, our
effective chiral Lagrangian is consistent with the known low-energy theorems
(Gell-Mann-Okubo and Gell-Mann-Oakes-Renner relations, partial conservation
of axial current (PCAC), Feynman-Hellmann relation between pion-nucleon
$\sigma$-term and the derivative of the nucleon mass, etc.).
The electromagnetic 
field is included into the effective Lagrangian (\ref{L_chir}) using the 
standard procedure, i.e. the interaction of quarks and charged mesons with 
photons is introduced using minimal substitution. 

To derive the properties of baryons, which are modeled as bound states 
of valence quarks surrounded by a meson cloud, we formulate perturbation 
theory and restrict the quark states to the ground-state contribution with
$\psi(x) = b_0 u_0(\vec{x})\exp(-i{\cal E}_0 t)$, 
where $b_0$ is the corresponding single-quark annihilation operator.
The quark wave function $u_0(\vec{x})$ belongs to the basis of potential
eigenstates used for expanding the quark field operator $\psi(\vec{x})$. 
In our calculation of matrix elements,  we project quark diagrams
on the respective baryon states. The baryon states are conventionally
set up by the product of the ${\rm SU(6)}$ spin-flavor and
${\rm SU(3)_c}$ color wave functions, where the nonrelativistic single
quark spin wave function is simply replaced by the relativistic solution
$u_0(\vec{x})$ of the Dirac equation
\begin{eqnarray}\label{Dirac_eq}
\left[ -i\gamma^0\vec{\gamma}\cdot\vec{\nabla} + \gamma^0 S(r) + V(r)
- {\cal E}_0 \right] u_0(\vec{x})=0,
\end{eqnarray}
where ${\cal E}_0$ is the single-quark ground-state energy.

For the description of baryon properties, we use the effective
potential $V_{\rm eff}(r)$ with a quadratic radial
dependence~\cite{Lyubovitskij:2001nm,Lyubovitskij:2000sf}:
\begin{eqnarray}\label{potential}
S(r) = M_1 + c_1 r^2, \hspace*{1cm} V(r) = M_2+ c_2 r^2
\end{eqnarray}
with the particular choice
\begin{eqnarray}
M_1 = \frac{1 \, - \, 3\rho^2}{2 \, \rho R} , \hspace*{1cm}
M_2 = {\cal E}_0 - \frac{1 \, + \, 3\rho^2}{2 \, \rho R} , \hspace*{1cm}
c_1 \equiv c_2 =  \frac{\rho}{2R^3} .
\end{eqnarray}
Here, $R$ and $\rho$ are parameters related to the ground-state quark wave
function $u_0$:
\begin{eqnarray}\label{eigenstate}
u_0(\vec{x};i) \, = \, N_0 \, \exp\biggl[-\frac{\vec{x}^{\, 2}}{2R^2}\biggr]
\, \left(
\begin{array}{c}
1\\
i \rho \, \vec{\sigma}(i)\cdot\vec{x}/R\\
\end{array}
\right)
\, \chi_s(i) \, \chi_f(i) \, \chi_c(i)\,,
\end{eqnarray}
where $N_0=[\pi^{3/2} R^3 (1+3\rho^2/2)]^{-1/2}$ is a normalization
constant; $\chi_s$, $\chi_f$, $\chi_c$ are the spin, flavor and color
quark wave functions, respectively. The index $"i"$ stands for the $i$-th
quark. The constant part of the scalar potential $M_1$ can be interpreted as
the constituent mass of the quark, which is simply the displacement of the
current quark mass due to the potential $S(r)$. The parameter $\rho$ is
related to the axial charge $g_A$ of the nucleon calculated in zeroth-order
(or 3q-core) approximation:
\begin{eqnarray}
g_A=\frac{5}{3}\biggl(1 - \frac{2\rho^2}{1+\frac{3}{2}\rho^2}\biggr)\,.
\end{eqnarray}
Therefore, $\rho$ can be replaced by $g_A$ using the matching condition (6).
The parameter $R$ is related to the charge radius
of the proton in the zeroth-order approximation as
\begin{eqnarray}
\la r^2_E \ra^P_{\rm LO} = \int d^3 x \, u^\dagger_0 (\vec{x}) \,
\vec{x}^{\, 2} \, u_0(\vec{x}) \, = \, \frac{3R^2}{2} \,
\frac{1 \, + \, \frac{5}{2} \, \rho^2}{1 \, + \, \frac{3}{2} \, \rho^2}.
\end{eqnarray}
In our calculations we use the value $g_A$=1.25. Therefore, we have only
one free parameter in our model, that is $R$ or  
$\la r^2_E \ra^p_{\rm LO}$. 
In previous publications $R$ was varied in the region from 0.55 fm to 0.65 fm,
which corresponds to a change of 
$\la r^2_E \ra^p_{\rm LO}$ from 0.5 to 0.7 fm$^2$. 
Note that for the given form of the effective potential~(\ref{potential})
the Dirac equation~(\ref{Dirac_eq}) can be solved analytically [for
the ground state see Eq.(\ref{eigenstate}), for excited states
see Ref.~\cite{Cheedket:2002ik}]. 

The expectation value of an operator $\hat A$ is then set up as:
\begin{equation}\label{hatA}
\hspace*{-1.5cm}\la \hat A \ra = {}^B\la \phi_0 |\sum^{\infty}_{n=1}
\frac{i^n}{n!}\int d^4 x_1 \ldots \int d^4 x_n T[{\cal L}_I (x_1)
\ldots{\cal L}_I (x_n) \hat A]|\phi_0 \ra^B_c,
\end{equation}
where the state vector $|\phi_0>$ corresponds to the unperturbed
three-quark state ($3q$-core). Superscript $``B"$ in the equation indicates
that the matrix elements have to be projected onto the respective
baryon states, whereas subscript $``c"$ refers to contributions from
connected graphs only. Here ${\cal L}_I (x)$ is the appropriate interaction 
Lagrangian. For the purpose of the present paper, we 
include in ${\cal L}_I (x)$ the linearized coupling of pseudoscalar fields 
with quarks and the corresponding coupling of quarks and mesons to the
electromagnetic field (see details in Ref.~\cite{Lyubovitskij:2001nm}): 
\eq\label{L_int}
{\cal L}_I (x) &=& - \bar \psi(x) i \gamma^5 \frac{\hat \Phi(x)}{F} 
S(r) \psi(x) - e A_\mu(x) \bar\psi(x) \gamma^\mu Q \psi(x) \nonumber\\
&-& e A_\mu(x) \sum\limits_{i,j=1}^{8}\biggl[f_{3ij} + 
\frac{f_{8ij}}{\sqrt{3}}\biggr] \Phi_i(x) \partial^\mu \Phi_j(x) 
 + \cdots \,, 
\en 
where $f_{ijk}$ are the SU(3) antisymmetric structure constants. 

For the evaluation of Eq.~(\ref{hatA}) we apply Wick's theorem
with the appropriate propagators for the quarks and pions.
For the quark propagator we use the vacuum Feynman propagator for a fermion
in a binding potential restricted to the ground-state quark wave function 
with 
\begin{eqnarray}\label{quark_propagator}
iG_0(x,y) = u_0(\vec{x}) \, \bar u_0(\vec{y}) \,
e^{-i{\cal E}_0 (x_0-y_0)} \, \theta(x_0-y_0)\,.
\end{eqnarray}
For the meson field we use the free Feynman propagator for a boson field with
\begin{eqnarray}
i\Delta_{ij}(x-y) = \la 0|T\{\Phi_i(x)\Phi_j(y)\}|0 \ra =
\delta_{ij}\int\frac{d^4k}{(2\pi)^4i} \, e^{-ik(x-y)} \,
\Delta_{\Phi}(k) \,,
\end{eqnarray}
where $\Delta_{\Phi}(k) = [M_\Phi^2-k^2-i0^+]^{-1}$ is the meson 
propagator in momentum space and $M_\Phi$ is the meson mass. 

The physical nucleon mass at one loop is given by 
\eq\label{nucleon mass}
m_N = m^{core}_N + \Delta m_N
\en 
where 
\eq\label{3q_core}
m^{core}_N = 
3 \bigg\{ {\cal E}_0 + \gamma \hat m  \biggr\} =  
3 \bigg\{ {\cal E}_0 + \frac{\gamma}{2B}M^2_\pi \biggr\}  
\en 
is the contribution of the three-quark core (the second term 
in the r.h.s. of Eq.~(\ref{3q_core}) is the contribution 
of the current quark mass) 
and 
\eq 
\Delta m_N = \Pi^{\rm MC} + \Pi^{\rm ME}
\en
is the nucleon mass shift due to the meson cloud contribution. 
The diagrams that contribute to the nucleon mass shift 
$\Delta m_N$ at one loop are shown in Fig.1 (see details in 
Refs.~\cite{Lyubovitskij:2001nm,Lyubovitskij:2000sf}). 
Fig.1a corresponds to the so-called meson-cloud (MC) contribution 
and Fig.1b is the meson-exchange (ME) contribution.  
The operators $\Pi^{\rm MC}$ and $\Pi^{\rm ME}$ are functions 
of the meson masses and are expressed in terms of the universal 
self-energy operator 
\eq 
\Pi(M_\Phi^2) \, = \, -I^{24}_{\Phi}, \,\,\,\, \Phi=\pi, K, \eta \,.  
\en
Here we introduce a notation for the structure integral in 
terms of which all further formulas can be expressed: 
\eq
I^{MN}_{\Phi} \, = \, \biggl(\frac{g_A}{\pi F}\biggr)^2 \int\limits_0^\infty 
\frac{dp \, p^N \, F_{\pi NN}^2(p^2)}{(M^2_{\Phi}+p^2)^{\frac{M}{2}}}, 
\,\,\,\, M, N=0,1,2,... 
\en
and 
\eq
I^{0N}_{\Phi} \, \equiv \,I_N \, &\equiv& \, \biggl(\frac{g_A}{\pi F}\biggr)^2 
\int\limits_0^\infty dp \, p^N \, F_{\pi NN}^2(p^2) = 
\biggl(\frac{g_A}{\pi F}\biggr)^2 
\biggl(\frac{2}{R^2}\biggr)^{\frac{N+1}{2}} \, \nonumber\\
&\times& \Gamma\biggl(\frac{N+1}{2}\biggr) 
\biggl( \frac{1}{2} - \frac{N+1}{4} \beta 
+ \frac{(N+1)(N+3)}{32} \beta^2 \biggr) \, , 
\en 
where $\beta = 2\rho^2/(2 - \rho^2) \,.$ 
The function $F_{\pi NN} (p^2)$ is the $\pi NN$ form factor 
normalized to unity at zero recoil ($p^2=0$):
\eq 
F_{\pi NN}(Q^2) = \exp\biggl(-\frac{Q^2R^2}{4}\biggr) \biggl\{ 1 \, - \, 
\frac{Q^2R^2}{4}\,\beta \biggr\}.  
\en
The meson cloud contributions to the mass shift are then given

1) in SU(2) as 
\eq 
\Pi^{\rm MC} &=& \frac{81}{400} \, \Pi(M_\pi^2) \,, \nonumber\\
\Pi^{\rm ME} &=& \frac{90}{400} \, \Pi(M_\pi^2) \,,
\en 

2) in SU(3) as 
\eq 
\Pi^{\rm MC} &=& \frac{81}{400} \, \Pi(M_\pi^2) 
           \, + \, \frac{54}{400} \, \Pi(M_K^2) 
           \, + \, \frac{9}{400} \, \Pi(M_\eta^2)\,, \nonumber\\
\Pi^{\rm ME} &=&   \frac{90}{400} \, \Pi(M_\pi^2) 
           \, - \, \frac{6}{400} \, \Pi(M_\eta^2)\,.
\en 
Exact expressions for the nucleon electromagnetic form factors can be 
found in Ref.~\cite{Lyubovitskij:2001nm}. Here we just present the typical 
results for the magnetic moments in SU(2) and SU(3). The two-flavor result is 
obtained from the three-flavor one when neglecting kaon and $\eta$-meson 
contributions. The magnetic moments of the nucleons, $\mu_p$ and $\mu_n$, 
are given by the expressions
\begin{eqnarray}\label{nucleon-magnetic-moment}
\hspace*{-1cm} \mu_p = {\mu^{LO}_p} \, \biggl[1+\delta-\frac{1}{400} 
\biggl\{ 26I^{34}_{\pi}+16I^{34}_{K}+4I^{34}_{\eta} \biggl\}\biggr]
+ \frac{m_N}{50} \biggl\{11I^{44}_{\pi}+I^{44}_{K}\biggr\}, \nonumber\\
\hspace*{-1cm} \mu_n = -\frac{2}{3}{\mu^{LO}_p} 
\,\biggl[1+\delta-\frac{1}{400} 
\biggl\{ 21I^{34}_{\pi}+21I^{34}_{K}+4I^{34}_{\eta} \biggl\}\biggr] 
- \frac{m_N}{50} \biggl\{11I^{44}_{\pi}+I^{44}_{K}\biggr\} \; ,
\end{eqnarray}
where
\eq
\mu^{LO}_p = \frac{2m_N\rho R}{1+\frac{3}{2}\rho^2}
\en
is the leading-order contribution to the proton magnetic moment. The factor
\eq
\delta = -\biggl({\hat m}+\frac{\Pi^{\rm MC}}{3}
\cdot \frac{1+\frac{3}{2}\rho^2}
{1-\frac{3}{2}\rho^2} \biggr) \, 
\frac{2-\frac{3}{2}\rho^2}{\biggl(1+\frac{3}{2}
\rho^2\biggr)^2} R\rho 
\en
defines the NLO correction to the nucleon magnetic moments due to 
the modification of the quark wave function~\cite{Lyubovitskij:2001nm}). 

\section{NUMERICAL RESULTS}

In this section we discuss the numerical results for the dependence of 
nucleon properties on a variation of the pseudoscalar meson masses. 

In Table 1 we present our results for the nucleon mass in dependence 
on the pion mass. Both the SU(2) version, considering only 
the pion cloud contribution, and the SU(3) variant, including in addition 
kaon and $\eta$-meson cloud contributions, are indicated.
The total result is normalized to the physical value (coinciding with 
the proton mass treated as the reference point) $m_N \equiv m_p = 938.27$ MeV 
by fixing the ground-state quark energy to ${\cal E}_0 \simeq 397$ MeV 
[in case of SU(2)] and ${\cal E}_0 \simeq 411$ MeV [in case of SU(3)]. 
We also indicate the separate contributions of the 3q-core and the meson 
cloud, and, in addition, the value obtained in the chiral limit, consistent 
with the values of Refs.~\cite{Faessler:2005gd,Borasoy:1996bx,Frink:2005ru}. 
For the dependence on the pion mass we choose mass values in the range of
$M_\pi^2 \simeq 0.15 - 1.2$ GeV$^2$ and the resulting nucleon mass is directly 
compared to either chiral extrapolations 
or lattice data~\cite{Orth:2005kq,Ali Khan:2003cu}. 
In both comparisons our results are consistent with the corresponding
values of either the
extrapolations or the lattice data. In Figs. 3 and 4 we indicate the full 
functional dependence of the nucleon mass on $M_\pi^2$, $M_K^2$ and 
$M_\eta^2$, respectively, and compare them (in case of the SU(2) 
$M_\pi^2$-dependence) to results of lattice QCD 
at order $p^3$ and $p^4$ of Ref.\cite{Procura:2006bj}. 
In Fig. 5 we compare the results for the  nucleon mass both in SU(2) and 
SU(3) to lattice QCD data from various 
collaborations~\cite{Orth:2005kq,Ali Khan:2003cu} 
as functions of $M_\pi^2$. 

In Table 2 we present our results for the nucleon magnetic moments 
in the SU(2) version for different values of the model scale parameter 
$R$ at $M_\pi = 0$ and at the physical pion mass $M^{\rm phys}_\pi$. 
In Table 3 we give the analogous results for SU(3).  
In Figs.6 and 7 we  draw the curves for the nucleon magnetic moments 
as functions of $M_\pi^2$ and compare them to results of lattice 
QCD~\cite{Holstein:2005db,Wang:2007iw}. Note, that the nucleon 
magnetic moments are not sensitive to a variation of the strange current 
quark mass $m_s$. In Figs. 8 and 9 we demonstrate the sensitivity of 
the nucleon magnetic moments as 
functions of $M_\pi^2$ on the variation of the scale parameter $R$. 

Finally, in Figs.10-12 we present results for the nucleon form factors 
$G_E^p(Q^2)$, $G_M^p(Q^2)/\mu_p$ and $G_M^n(Q^2)/\mu_n$ at different values 
of $M_\pi^2$. A larger pion mass leads to an increase of the normalized 
(at $Q^2 = 0$) form factors.

\section{CONCLUSIONS}

In this work we apply the perturbative chiral quark model at one loop
to describe the dependence of nucleon properties on the meson masses.
It has been 
previously verified that the model is successful in the explanation of many 
aspects of nucleon properties~[23-25], such as magnetic moments, the axial 
vector form factor, the $N \to \Delta$ transition amplitude, the meson-nucleon 
sigma-term and $\pi N$ nucleon scattering. Here we demonstrate
that the meson mass dependence
of nucleon properties such as the mass,  
magnetic moments, electromagnetic form factors, both for the two
and three flavor variants, is reasonably described in comparison
to present lattice data and extrapolations 
of these results. Given also the simplicity of this model approach,
the evaluation at one loop seems sufficient to correctly describe
the pion mass dependence of the discussed observables.

\section*{Acknowledgments}

This work was supported by the DFG under contracts FA67/31-1 and
GRK683. This research is also part of the EU Integrated
Infrastructure Initiative Hadronphysics project under contract
number RII3-CT-2004-506078 and President grant of Russia
"Scientific Schools"  No. 5103.2006.2. The stay in T\"ubingen of 
Chalump Oonariya was supported by the DAAD under PKZ:A/05/57631.

%%%%%%%%%%%%%%%%%
%   TABLES      %
%%%%%%%%%%%%%%%%%

\begin{table}[htbp]
\caption{Nucleon mass (in GeV) in the SU(2) and SU(3) versions and 
at different values for $M_\pi ^2$}
\begin{center}
\def\arraystretch{0.85}
\begin{tabular}{ l | c | c | l } \hline\hline
Nucleon mass    &  SU(2)     & SU(3)     & Other approaches\\ \hline\hline
3q-core         & 1.203      & 1.247     & \hspace{1cm}-\\
Meson loops (total)& -0.265  & -0.309    & \hspace{1cm}-\\
$\pi$ loops     & -0.265     & -0.265    & \hspace{1cm}-\\
$K$ loops       & -          & -0.042    & \hspace{1cm}-\\
$\eta$ loops    & -          & -0.002    & \hspace{1cm}-\\ 
Total           & 0.93827    & 0.93827   & \hspace{1cm}-\\ \hline \hline
Chiral limit    & 0.887      & 0.831     
                & 0.880~\cite{Procura:2003ig,Bernard:2003rp}; 
                                           0.883~\cite{Procura:2006bj};\\ 
                &            &           & 0.770(110)~\cite{Borasoy:1996bx}; 
		                           0.890(180)~\cite{Frink:2005ru};\\ 
                &            &           & 0.832~\cite{Faessler:2005gd}\\
\hline 
$M^2_\pi$ (in GeV$^2$) & & & \\
\hline 
0.153 & 1.122 & 1.128 & 1.182(26)~\cite{Orth:2005kq}\\ 
0.162 & 1.133 & 1.138 & 1.195(42)~\cite{Orth:2005kq}\\ 
0.175 & 1.148 & 1.154 & 1.104(20)~\cite{Orth:2005kq}\\ 
0.240 & 1.212 & 1.220 & 1.228(31)~\cite{Orth:2005kq}\\ 
0.348 & 1.310 & 1.320 & 1.356(21)~\cite{Orth:2005kq}\\
0.413 & 1.360 & 1.371 & 1.377(19)~\cite{Orth:2005kq}\\
0.462 & 1.400 & 1.412 & 1.410(17)~\cite{Orth:2005kq}\\
0.557 & 1.475 & 1.490 & 1.533(28)~\cite{Orth:2005kq}\\ 
0.588 & 1.500 & 1.515 & 1.509(16)~\cite{Orth:2005kq}\\
0.678 & 1.566 & 1.582 & 1.637(27)~\cite{Orth:2005kq}\\
0.774 & 1.638 & 1.656 & 1.631(30)~\cite{Orth:2005kq}\\
0.810 & 1.664 & 1.682 & 1.619(16)~\cite{Orth:2005kq}\\
\hline 
0.258 & 1.231 & 1.239 & 1.253(15)~\cite{Ali Khan:2003cu}\\ 
0.271 & 1.240 & 1.248 & 1.275(82)~\cite{Ali Khan:2003cu}\\ 
0.297 & 1.266 & 1.275 & 1.300(22)~\cite{Ali Khan:2003cu}\\ 
0.310 & 1.275 & 1.284 & 1.320(19)~\cite{Ali Khan:2003cu}\\ 
0.314 & 1.280 & 1.288 & 1.412(61)~\cite{Ali Khan:2003cu}\\ 
0.354 & 1.314 & 1.324 & 1.348(13)~\cite{Ali Khan:2003cu}\\
0.502 & 1.431 & 1.444 & 1.497(77)~\cite{Ali Khan:2003cu}\\
0.514 & 1.442 & 1.456 & 1.506(94)~\cite{Ali Khan:2003cu}\\
0.536 & 1.458 & 1.742 & 1.509(18)~\cite{Ali Khan:2003cu}\\
0.540 & 1.462 & 1.476 & 1.519(11)~\cite{Ali Khan:2003cu}\\
0.578 & 1.493 & 1.507 & 1.657(26)~\cite{Ali Khan:2003cu}\\
0.607 & 1.515 & 1.530 & 1.629(20)~\cite{Ali Khan:2003cu}\\
0.776 & 1.640 & 1.658 & 1.679(36)~\cite{Ali Khan:2003cu}\\
0.874 & 1.710 & 1.730 & 1.741(29)~\cite{Ali Khan:2003cu}\\
0.883 & 1.717 & 1.736 & 1.781(15)~\cite{Ali Khan:2003cu}\\ 
0.894 & 1.725 & 1.744 & 1.878(28)~\cite{Ali Khan:2003cu}\\ 
0.901 & 1.730 & 1.749 & 1.785(35)~\cite{Ali Khan:2003cu}\\ 
0.913 & 1.738 & 1.758 & 1.798(85)~\cite{Ali Khan:2003cu}\\
0.921 & 1.744 & 1.764 & 1.801(14)~\cite{Ali Khan:2003cu}\\
0.940 & 1.758 & 1.778 & 1.809(17)~\cite{Ali Khan:2003cu}\\ 
1.201 & 1.943 & 1.965 & 2.063(26)~\cite{Ali Khan:2003cu}\\ 
\hline\hline
\end{tabular}
\end{center}
\end{table}

\begin{table} 
\caption{Nucleon magnetic moments in SU(2)}
\begin{center}
\begin{tabular}{ c | c | c | c | c } \hline\hline
\multicolumn{1}{c|}{}
&\multicolumn{2}{|c|}{\bf $\mu_p$}
&\multicolumn{2}{|c}{\bf $\mu_n$} \\ \cline{2-3} \cline{4-5}
R (fm) &$M_\pi=0$ & $M^{\rm phys}_\pi$ & $M_\pi=0$ & $M^{\rm phys}_\pi$\\ 
\hline\hline
0.50   & 3.896  & 2.984 & -2.948 & -2.015  \\
0.55   & 3.828  & 2.947 & -2.871 & -1.970  \\
0.60   & 3.796  & 2.945 & -2.823 & -1.952  \\
0.65   & 3.792  & 2.969 & -2.797 & -1.954  \\
0.70   & 3.803  & 3.012 & -2.789 & -1.973  \\ \hline\hline
\end{tabular}
\end{center}

\vspace*{2cm} 

\caption{Nucleon magnetic moments in SU(3) for fixed
masses $M^2_K$ and $M^2_\eta$}
\begin{center}
\begin{tabular}{ c | c | c | c | c } \hline\hline
\multicolumn{1}{c|}{} 
&\multicolumn{2}{|c|}{\bf $\mu_p$}
&\multicolumn{2}{|c}{\bf $\mu_n$} \\ \cline{2-3} \cline{4-5}
R (fm) &$M_\pi=0$ & $M^{\rm phys}_\pi$ & $M_\pi=0$ & $M^{\rm phys}_\pi$\\ 
\hline\hline
0.50   & 3.512  & 2.598 & -2.976 & -2.043  \\
0.55   & 3.466  & 2.585 & -2.984 & -1.993  \\
0.60   & 3.456  & 2.605 & -2.843 & -1.972  \\
0.65   & 3.471  & 2.648 & -2.815 & -1.972  \\
0.70   & 3.505  & 2.709 & -2.804 & -1.988  \\ 
\hline\hline
\end{tabular}
\end{center}
\end{table}

\newpage
%%%%%%%%%%%%%%%%%
%   GRAPHICS    %
%%%%%%%%%%%%%%%%%

\begin{figure}
\centering{\
\epsfig{figure=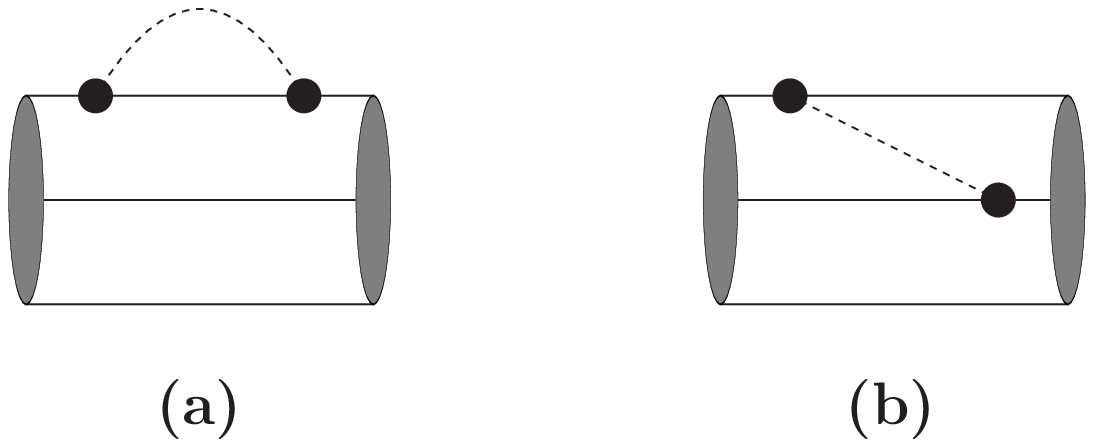,scale=0.75}} 
\caption{Diagrams contributing to the nucleon mass shift.}
\vspace*{2cm}
\centering{\
\epsfig{figure=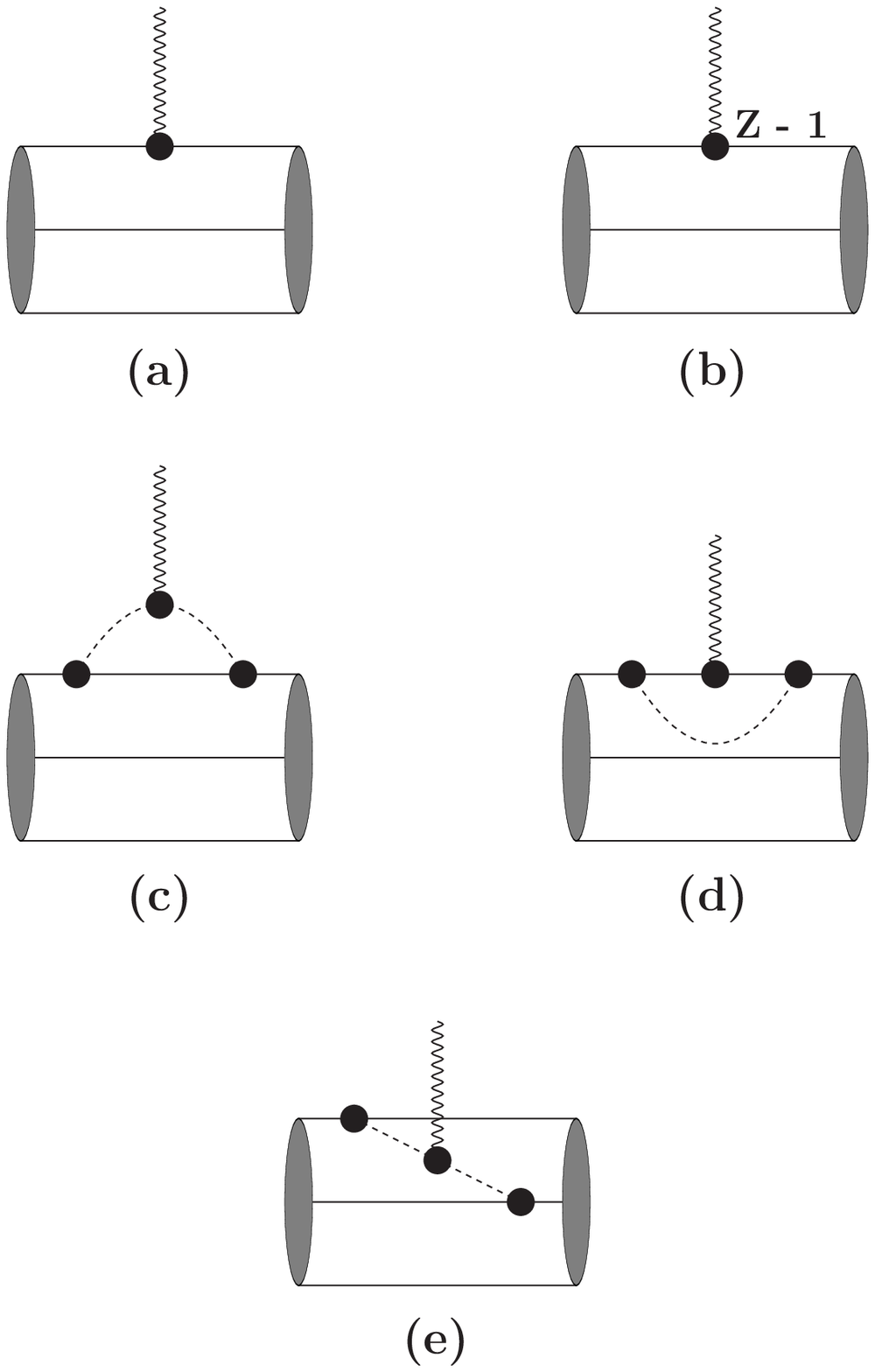,scale=0.75}} 
\caption{Diagrams contributing to the nucleon electromagnetic form factors.}
\end{figure}

\begin{figure}[htbp]
\vspace*{3.8cm}
\centering{\
\epsfig{figure=fig3.eps,scale=0.38}} 
\caption{Dependence of nucleon mass $m_N$ on meson mass $M^2_\Phi$:
$m_N(M^2_\pi)$ in SU(2) (dotted line) and 
$m_N(M^2_\pi)$, $m_N(M_K^2)$ $m_N(M_\eta^2)$ 
in SU(3) (the other lines).}
\vspace*{3.5cm}
\centering{\
\epsfig{figure=fig4.eps,scale=0.38}} 
\caption{Dependence of nucleon mass $m_N$ on pion mass $M^2_\pi$:
$m_N(M^2_\pi)$ in SU(2) (dotted line), 
SU(3) (solid line) and from lattice 
QCD~\cite{Procura:2006bj} (the others) at order $p^3$ and~$p^4$. }
\end{figure} 

\begin{figure}[htbp]
\vspace*{6.8cm}
\centering{\
\epsfig{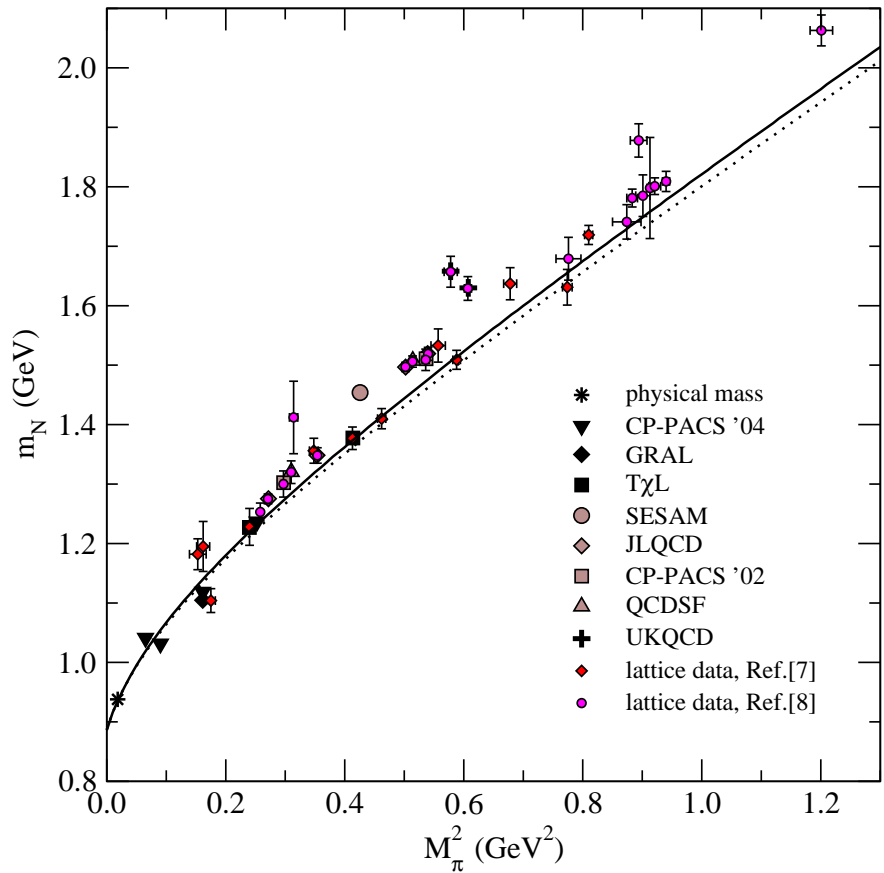}}
\caption{Nucleon dependencies $m_N(M^2_\pi)$ in SU(2) (dotted line) and 
in SU(3) (solid line) are compared to
data from various collaborations as a function of $M^2_{\pi}$. }
\end{figure}

\begin{figure}[htbp]
\vspace*{3.5cm}
\centering{\
\epsfig{figure=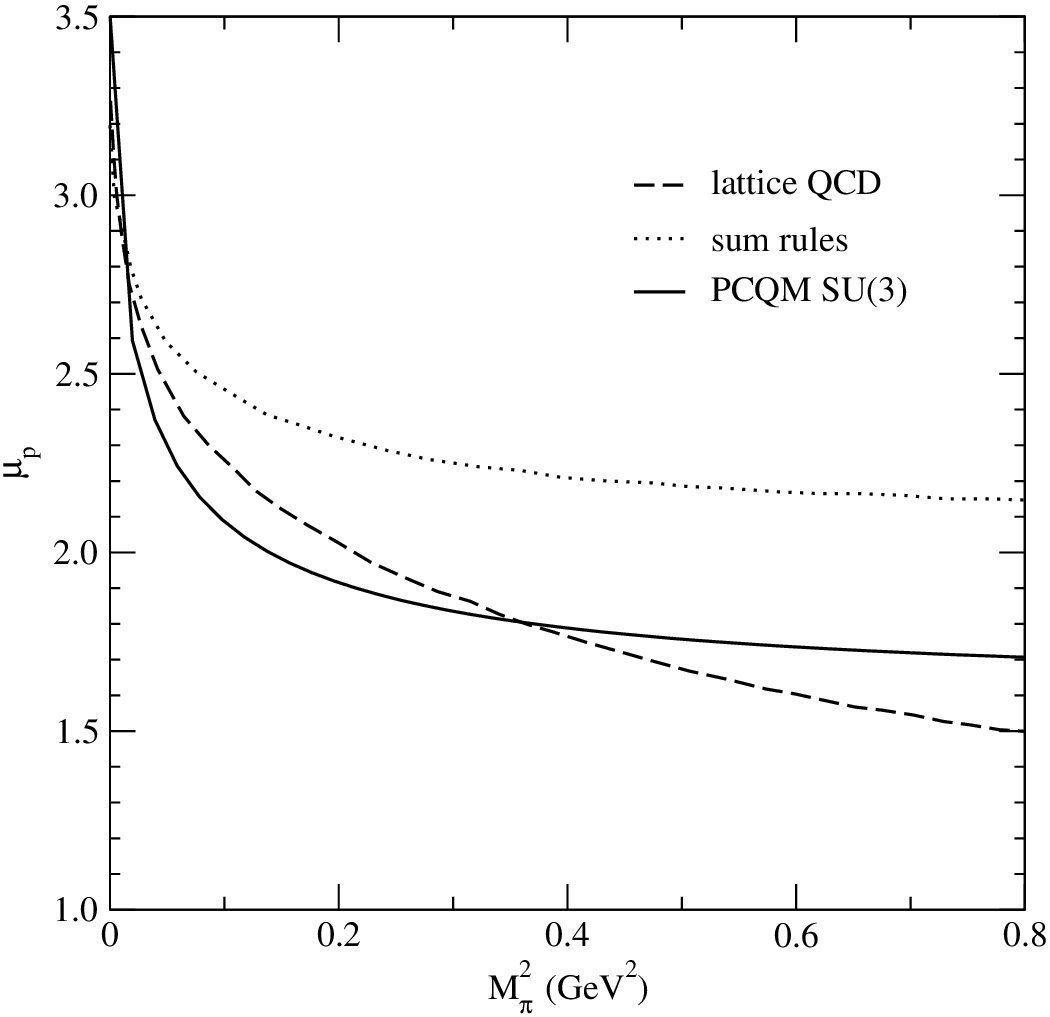,scale=0.37}}
\caption{$M^2_\pi$-dependence of the proton magnetic moment
$\mu_p(M^2_{\pi})$ to one-loop from 
sum rules~\cite{Holstein:2005db} (dotted curve), 
lattice QCD~\cite{Wang:2007iw} (dashed curve) 
and our results in SU(3) (solid curve).}  
\vspace*{3.5cm}
\centering{\
\epsfig{figure=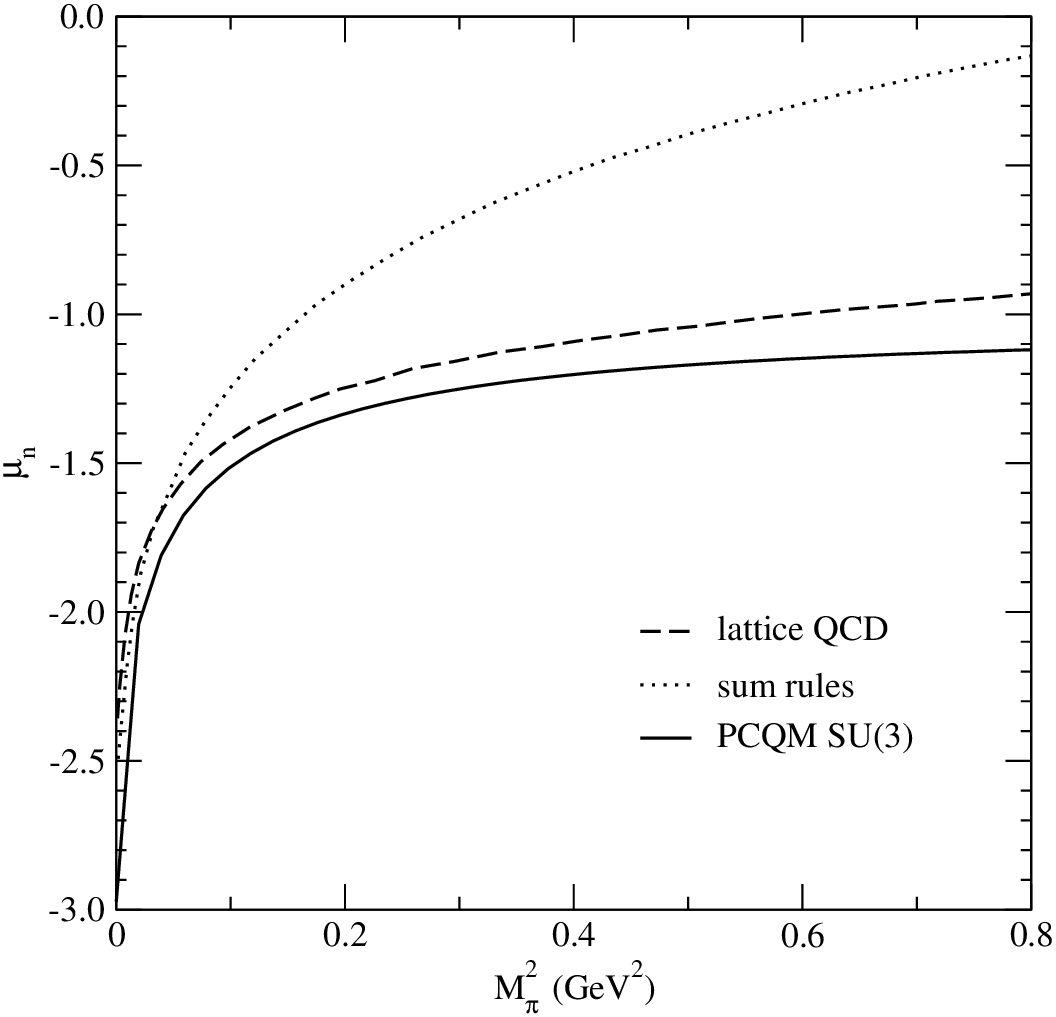,scale=0.37}}
\caption{$M^2_\pi$-dependence of the neutron magnetic moment
$\mu_n(M^2_{\pi})$ to one-loop from sum 
rules~\cite{Holstein:2005db} (dotted curve), 
lattice QCD~\cite{Wang:2007iw} (dashed curve) and 
our results in SU(3) (solid curve).}  
\end{figure}

\begin{figure}[htbp]
\vspace*{3.5cm}
$\begin{array}{c@{\hspace*{2cm}}c}
\multicolumn{1}{l}{\mbox{}} & \multicolumn{1}{l}{\mbox{}}\\ [-0.53cm]
\epsfig{file=fig8a.eps, scale=0.38} & 
\epsfig{file=fig8b.eps, scale=0.38} \\ 
\mbox{(a)} & \mbox{(b)}
\end{array}$
\vspace*{-.3cm}
\caption{$\mu_N$ at $M^2_\pi= 0 - 1$ GeV$^2$ and 
$R = 0.5 - 0.7$ fm in SU(2).}

\vspace*{3.5cm}

\begin{center}
$\begin{array}{c@{\hspace*{2cm}}c}
\multicolumn{1}{l}{\mbox{}} & \multicolumn{1}{l}{\mbox{}}\\ [-0.53cm]
\epsfig{file=fig9a.eps, scale=0.38} & 
\epsfig{file=fig9b.eps, scale=0.38} \\ 
\mbox{(a)} & \mbox{(b)}
\end{array}$
\end{center}
\vspace*{-.3cm}
\caption{$\mu_N$ at $M^2_\pi= 0 - 1$ GeV$^2$ and 
$R = 0.5 - 0.7$ fm in SU(3).}
\end{figure}

\begin{figure}[htbp]
\vspace*{3.8cm}
\centering{\
\epsfig{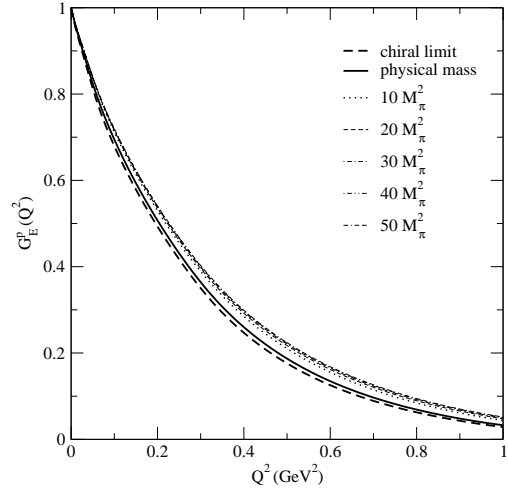}}
\caption{Proton charge form factor $G_E^p (Q^2)$ 
as function of $M_\pi^2$ in SU(2).}
\end{figure}

\begin{figure}[htbp]
\vspace*{3.8cm}
\centering{\
\epsfig{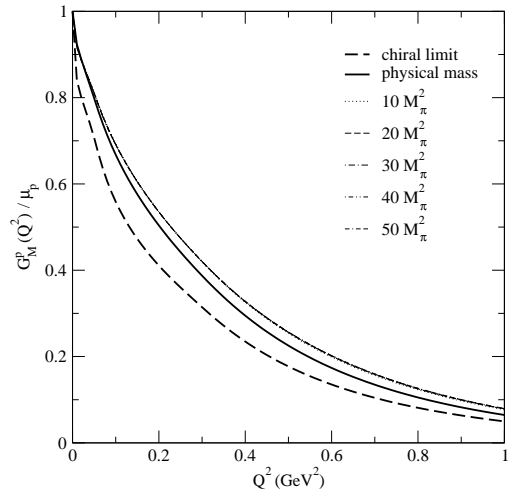}}
\caption{Proton magnetic form factor $G_M^p (Q^2)$ 
as function of $M_\pi^2$ in SU(2).}
\end{figure}

\begin{figure}[htbp]
\vspace*{3.8cm}
\centering{\
\epsfig{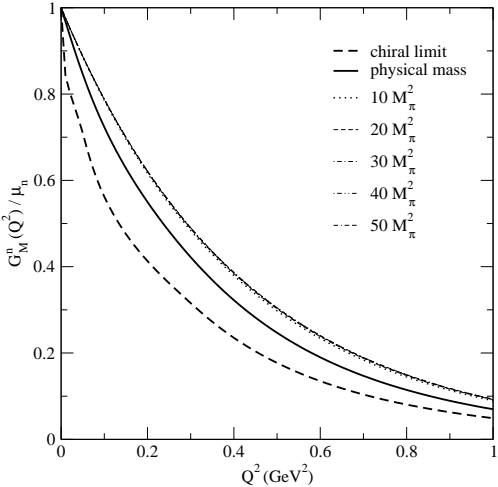}}
\caption{Neutron magnetic form factor $G_M^n (Q^2)$ 
as function of $M_\pi^2$ in SU(2).} 
\end{figure}

\end{document}